\documentclass{article}

\usepackage[dblblindworkshop,final]{neurips_2026}

\usepackage[utf8]{inputenc}
\usepackage[T1]{fontenc}
\usepackage{hyperref}
\hypersetup{hidelinks}
\usepackage{url}
\usepackage{booktabs}
\usepackage{amsfonts}
\usepackage{amsmath}
\usepackage{amssymb}
\usepackage{mathtools}
\usepackage{nicefrac}
\usepackage{microtype}
\usepackage{xcolor}
\usepackage{graphicx}
\usepackage{multirow}
\usepackage{array}
\usepackage{tikz}
\usepackage{caption}
\usepackage{subcaption}
\usetikzlibrary{arrows.meta,positioning,fit,calc}

\newcommand{\R}{\mathbb{R}}

\title{Physics Foundation Models for Efficient \\ Control of Fluid Flows}

\author{
  Mahindra Rautela$^{1}$ \qquad
  Shaifalee Saxena$^{1,2}$ \qquad
  Alexander Scheinker$^{1}$ \\[0.5em]
  $^{1}$Los Alamos National Laboratory, Los Alamos, NM, USA \\
  $^{2}$University of New Mexico, Albuquerque, NM, USA \\[0.3em]
  \texttt{\{mrautela,shaifalees,ascheink\}@lanl.gov}
}
\begin{document}
\maketitle

\begin{abstract}
We investigate whether a physics foundation model (PFM) pretrained for PDE forecasting can be repurposed for closed-loop fluid-flow control. We construct a PFM-based controller by attaching lightweight Gaussian actor and critic heads to frozen pretrained representations and train the policy using proximal policy optimization (PPO). We evaluate the approach on HydroGym's two-dimensional Kolmogorov-flow environment using vorticity observations at \(64\times64\) and \(128\times128\) resolutions and Reynolds numbers \(\mathrm{Re}=100\) and \(200\). Across these settings, we compare the frozen PFM controller against a baseline MLP policy trained with the same PPO framework. The PFM controller achieves consistently higher reward while requiring smaller executed action magnitudes over training, indicating improved control performance with reduced control effort. These results provide initial evidence that representations learned through self-supervised next-frame prediction can be transferred directly to closed-loop fluid flow control. \noindent\textbf{Project page:} \url{https://mahindrautela.github.io/morph}.

\end{abstract}

\section{Introduction}

Learning to predict the evolution of partial differential equations (PDEs) has progressed from task-specific surrogates toward pretrained models that reuse representations across physical systems. Recent physics foundation models are pretrained on heterogeneous collections of PDE trajectories and transfer to downstream physical systems with substantially less task-specific data than training a comparable model from scratch~\citep{hao2024dpot,herde2024poseidon,mccabe2025walrus,rautela2026morph}. 

Although physics foundation models have demonstrated strong transfer and sample efficiency, their applications have largely focused on forward prediction tasks, such as next-state prediction and autoregressive rollout~\citep{rautela2026out}. These capabilities are valuable for accelerating simulation, but they do not directly address closed-loop control, where a policy must map the current state $s_t$ to an action $a_t$ while optimizing long-horizon return under the system dynamics. Recent work has begun to extend PDE foundation models to inverse estimation~\citep{rautela2026inverse}, yet their use as closed-loop \emph{actors} remains comparatively unexplored. This motivates the central question of this work: \emph{can a foundation model pretrained for PDE prediction be repurposed for effective fluid-flow control?}

HydroGym provides a timely testbed for investigating this question \citep{lagemann2026hydrogym}. It is a solver-independent reinforcement-learning platform comprising multiple validated flow-control environments. In this work, we focus on its two-dimensional chaotic Kolmogorov-flow environment characterized by extreme energy dissipation events from nonlinear energy transfer across scales. The environment provides full-field vorticity observations as the state at varying Reynolds numbers, while control is applied through four continuous forcing amplitudes.

\begin{figure}[t]
\centering
\includegraphics[width=\textwidth,trim=0cm 0cm 0cm 0cm, clip]{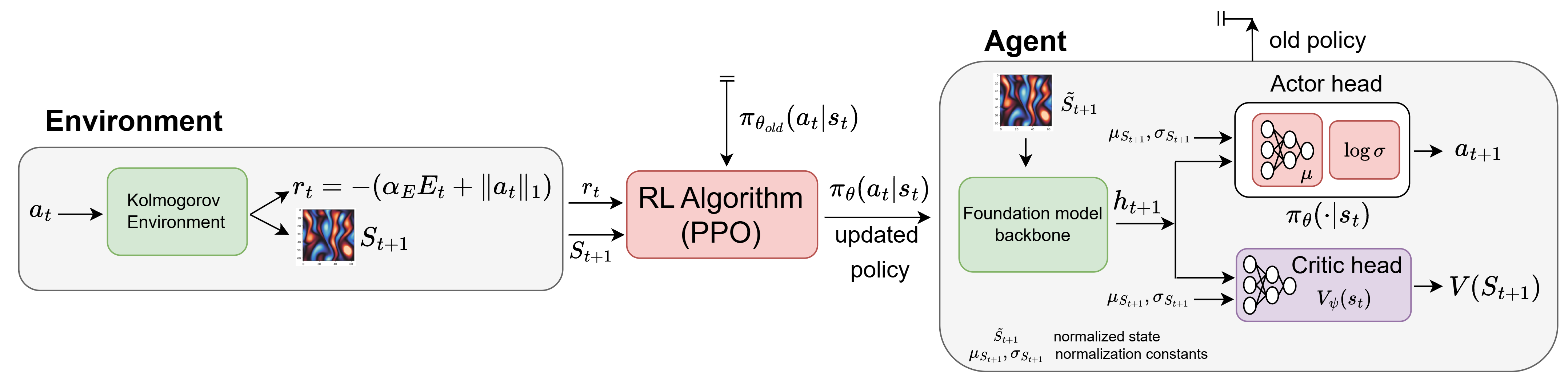}
\caption{\textbf{Repurposing a physics foundation model for closed-loop flow control.} The latent state representation is passed to lightweight actor and critic heads, which parameterize a Gaussian control policy and estimate the state value, respectively. PPO updates the control heads from closed-loop interaction with the Kolmogorov-flow environment, while the pretrained foundation-model backbone remains frozen.}
\label{fig:morph_ctrl}
\vspace{-4mm}
\end{figure}

In this paper, we repurpose a pretrained physics foundation model, MORPH-Ti, as a flow-control policy by removing its next-state decoder, pooling the learned spatial representations, and attaching lightweight actor and critic heads \citep{rautela2026morph}. Fig.~\ref{fig:morph_ctrl} summarizes the resulting prediction-to-control adaptation and closed-loop training framework. Table~\ref{tab:fm_selection} in  Appendix~\ref{app:setup} summarizes representative alternative foundation models and motivates our model selection.

Our main comparison deliberately keeps the pretrained MORPH backbone frozen and trains only the actor and critic heads. We compare this controller against a conventional PPO MLP that receives the same full-field vorticity observations and is trained under the same budget. To our knowledge, this is the first study to directly repurpose representations from a pretrained PDE foundation model as a closed-loop RL policy for fluid-flow control. 

\section{Related work} \label{related_works}

\paragraph{Learning PDE dynamics with surrogate models.}
Data-driven models for physical dynamics span continuous-time models, neural operators, convolutional architectures, and reduced-order latent representations. Neural ODEs learn continuous-time dynamics directly from data~\citep{chen2018neuralode}. DeepONet~\citep{lu2021deeponet} and the Fourier Neural Operator (FNO)~\citep{li2020fno} learn mappings between function spaces, while U-Net-style and U-shaped neural operators provide multiscale encoder-decoder alternatives~\citep{ronneberger2015unet,rahman2022uno}. Latent evolution models compress high-dimensional states and learn temporal evolution in a reduced representation; examples include convolutional autoencoder--recurrent models for advection-dominated systems~\citep{maulik2021reduced} and $\beta$-VAE--transformer models for chaotic fluid flows~\citep{solerarico2024beta}. These approaches are typically optimized for a single system or family of closely related systems.

\paragraph{PDE foundation models.}
PDE foundation models instead pretrain a shared backbone across heterogeneous dynamical systems. DPOT combines autoregressive denoising with a scalable Fourier-attention operator transformer~\citep{hao2024dpot}. Poseidon uses a multiscale operator transformer and semigroup-based training for transferable fluid-dynamics representations~\citep{herde2024poseidon}. PDE-Transformer develops a scalable general-purpose transformer for regular-grid physics simulations~\citep{holzschuh2025pdetransformer}. Walrus extends cross-domain pretraining across diverse continuum dynamics and emphasizes stable long-horizon forecasting~\citep{mccabe2025walrus}. MORPH introduces a modality-agnostic convolutional vision-transformer backbone that handles mixed fields, dimensions, and resolutions through component convolutions, inter-field cross-attention, and axial attention~\citep{rautela2026morph}. Although these models differ architecturally, their central downstream task remains predominantly prediction or reconstruction. Our work asks whether such pretrained representations can instead serve directly as the state encoder of an RL actor.

\paragraph{Reinforcement learning for flow control.}
Deep RL has become a powerful tool for active flow control. Early work demonstrated that neural policies can discover effective closed-loop strategies for two-dimensional cylinder wakes~\citep{rabault2019afc}, followed by experimental and simulation studies of bluff-body control~\citep{fan2020reinforcement}, sensor-efficient control~\citep{paris2021robust}, turbulent drag reduction~\citep{guastoni2023deep}, and increasingly complex three-dimensional and turbulent configurations~\citep{font2025deep,suarez2025flow}. Recent studies illustrate both the capability and cost of this paradigm: ~\cite{montala2026wing} train multi-agent PPO policies directly in three-dimensional wing CFD, while~\cite{mondal2026hypersonic} use off-policy DRL for robust hypersonic inlet control and identify scientific foundation models as a promising route toward more scalable learning-enabled control. HydroGym consolidates many such settings into a common benchmark and demonstrates that transfer can reduce the cost of learning new policies~\citep{lagemann2026hydrogym}. In contrast to prior transfer across flow conditions or policy environments, we transfer the \emph{representation itself}: a backbone pretrained to model PDE dynamics becomes the state encoder of the RL actor.

\section{From PDE forecasting to closed-loop control} \label{app:setup}

\subsection{Problem formulation}
We consider a Markov decision process with flow state $s_t\in\mathcal{S}$, continuous action $a_t\in\mathcal{A}$, reward $r_t$, and transition $s_{t+1}\sim\mathcal{P}(\cdot\mid s_t,a_t)$ induced by the numerical flow solver. The policy $\pi_\theta(a_t\mid s_t)$ is trained to maximize the discounted return
\begin{equation}
J(\theta)=\mathbb{E}_{\pi_\theta}\left[\sum_{t=0}^{T-1}\gamma^t r_t\right].
\end{equation}
A PDE foundation model is instead pretrained for a predictive map $F_\phi:s_t\mapsto \hat{s}_{t+1}$ (or a related temporal operator). We retain the pretrained state representation but replace the task-specific output map.

\subsection{Lightweight control adaptation}
Figure~\ref{fig:morph_ctrl} summarizes the adaptation. For MORPH, the released implementation produces a transformer representation $z\in\R^{B\times T\times N\times d}$ immediately before its physical-field decoder. We bypass that decoder and mean-pool the final-time tokens,
\begin{equation}
h_t=\frac{1}{N}\sum_{i=1}^{N}\mathrm{LN}\!\left(z_{t,i}\right),
\end{equation}
then attach separate actor and value heads that parameterize a diagonal-Gaussian policy and scalar value function. The exact head architecture and gradient routing are described below. At deployment only the backbone and actor are required.

For MORPH-Ti, a real-space $n\times n$ scalar field is represented in UPTF-7 form with $(B,T,F,C,D,H,W)=(B,1,1,1,1,n,n)$ for $n\in\{64,128\}$. In the reported MORPH runs, each vorticity frame is normalized using its own spatial mean and standard deviation before entering the backbone ($\mu_{S_{t+1}}, \sigma_{S_{t+1}}$ in Fig.~\ref{fig:morph_ctrl}). The original frame mean and standard deviation are retained and appended to the pooled latent representation before the actor and critic heads. We use the general pretrained MORPH-Ti checkpoint before any Kolmogorov-specific predictive fine-tuning so that control trajectories and rewards are unseen during pretraining, and the backbone is kept frozen during PPO training.

\begin{table}[h]
\centering
\small
\setlength{\tabcolsep}{4.8pt}
\caption{Kolmogorov-flow and PPO configuration used for the reported comparisons. The same control interval, reward, and PPO optimization settings are used across resolutions and Reynolds numbers.}
\label{tab:setup}
\begin{tabular}{ll@{\hspace{8mm}}ll}
\toprule
\textbf{Environment} & \textbf{Value} & \textbf{PPO} & \textbf{Value} \\
\midrule
Reynolds number & $\{100,200\}$ & Baseline / MORPH head LR & $10^{-4}$ \\
Grid & $64\times64,\;128\times128$ & Parallel envs & 8 \\
Domain & $[0,2\pi]^2$ & Rollout steps / update & 10 \\
$\Delta t$ & $10^{-3}$ & Minibatches & 8 \\
Control interval & 10.0 & Update epochs & 10 \\
Action modes & $4,5,6,7$ & Discount $\gamma$ & 0.99 \\
Action bounds & $[-0.5,0.5]^4$ & GAE $\lambda$ & 0.985 \\
Environment episode limit & 1000 & PPO clip & 0.2 \\
Training budget & 80,000 interactions & PPO updates & 1000 \\
State & $\omega_t\in\R^{n\times n}$ & Entropy coeff. & 0.01 \\
Energy weight $\alpha_E$ & -1 & Value coeff. & 0.5 \\
Action weight $\lambda_a$ & 1 & Max grad norm & 0.5 \\
Random seed & 30 & MORPH backbone & frozen \\
\bottomrule
\end{tabular}
\end{table}

\subsection{Actor--critic architectures}
\paragraph{MORPH actor--critic.}
The released MORPH-Ti backbone used here has an $8\times8$ spatial patch size, a 256-dimensional embedding, four transformer blocks, and four attention heads. A $64\times64$ field produces $8\times8=64$ spatial tokens, whereas a $128\times128$ field produces $16\times16=256$ tokens. The final-time tokens are layer-normalized and mean-pooled to obtain $h_t\in\mathbb{R}^{256}$. With frame normalization enabled, the raw spatial mean $\mu_{S_{t}}$ and standard deviation $\sigma_{S_{t}}$ are appended, giving $\tilde h_t=[h_t,\mu_{S_t},\sigma_{S_t}]\in\mathbb{R}^{258}$. The actor is a lightweight MLP: $258\rightarrow256\rightarrow4$ with a $\tanh$ hidden layer. Its output parameterizes the mean of a four-dimensional diagonal Gaussian policy with a learned state-independent log standard deviation. The critic has the corresponding structure $258\rightarrow256\rightarrow1$.

The critic is evaluated on a stop-gradient copy of the pooled representation, so the value loss trains the critic without propagating gradients into the MORPH representation. In the reported frozen-backbone experiments, no backbone parameters are updated by either the policy or value objective. PPO optimizes only the lightweight control heads and policy log standard deviation.

\paragraph{PPO baseline.}
The full-field PPO baseline receives the same physical $n\times n$ vorticity field as MORPH and flattens it before the policy network. Its actor and critic are separate MLPs, each with two 256-unit $\tanh$ hidden layers: $n^2\rightarrow256\rightarrow256\rightarrow4$ for the actor and $n^2\rightarrow256\rightarrow256\rightarrow1$ for the critic. Thus the input dimension is 4096 at $64\times64$ and 16384 at $128\times128$. The actor parameterizes a diagonal Gaussian policy with a learned state-independent log standard deviation. The two branches share no hidden parameters, so value-function gradients do not modify the actor representation.

\subsection{PPO motivation and optimization}

Our PPO implementation builds on the HydroGym training pipeline, which follows the PureJaxRL PPO implementation, with modifications for full-field vorticity observations, MORPH-based actor-critic policies, and numerical-failure handling. Because the released MORPH checkpoint is implemented in PyTorch while HydroGym uses JAX, we retain the flow solver in JAX and execute the MORPH policy and PPO updates in PyTorch, converting observations and actions between the two frameworks at the environment--policy interface.

All controllers are trained with proximal policy optimization (PPO)~\citep{schulman2017ppo}; the shared optimization settings are summarized in Table~\ref{tab:setup}. PPO alternates between collecting trajectories with the current policy and optimizing a clipped surrogate objective. For a collected action $a_t$, the likelihood ratio
\[
\rho_t(\theta)=\frac{\pi_\theta(a_t\mid s_t)}{\pi_{\theta_{\mathrm{old}}}(a_t\mid s_t)}
\]
is used in the policy loss
\[
\mathcal{L}_{\mathrm{policy}}
=-\mathbb{E}_t\!\left[
\min\!\left(
\rho_t\hat{A}_t,
\operatorname{clip}(\rho_t,1-\epsilon,1+\epsilon)\hat{A}_t
\right)
\right].
\]
Because PPO is on-policy, each update uses trajectories generated by the current controller rather than a replay buffer populated by older policies. Its clipped objective further discourages large changes in action probabilities within an update. This combination makes PPO a conservative choice for adapting a pretrained representation in our pilot study, although the clipping acts in policy space and does not directly constrain changes to the MORPH parameters. A related precedent exists in language-model RLHF: PPO was used to adapt a pretrained 774M-parameter GPT-2 model~\citep{ziegler2019fine} and later GPT-3-based InstructGPT models \citep{ouyang2022training}. More details about PPO are given in Appendix~\ref{app:ppo}.

\subsection{Controlled comparison}
A pretrained model can appear better than a conventional RL baseline for reasons unrelated to the representation itself. We therefore keep the physical observation, action space, reward, PPO update rule, and interaction budget fixed between the conventional full-field baseline and the MORPH controller. The key experimental choice is that the MORPH backbone remains frozen: only the actor/critic heads and policy log standard deviation are trained. The comparison therefore asks whether a representation learned entirely from PDE forecasting data can serve as a useful control encoder without downstream modification of the foundation model. Since the MLP and MORPH encoders differ architecturally, this comparison should be interpreted as a test of the usefulness of the frozen pretrained representation rather than a causal isolation of pretraining alone. We repeat the comparison across spatial resolutions and Reynolds numbers to test whether the observed advantage is specific to a single discretization or flow regime.

\section{Experiments}

\subsection{Kolmogorov-flow control in HydroGym}
HydroGym's Kolmogorov environment solves the two-dimensional incompressible Navier--Stokes equations in vorticity form on the periodic domain $[0,2\pi]^2$. With $\omega=\nabla\times u$ and Reynolds number $\mathrm{Re}\in\{100,200\}$, the controlled dynamics can be expressed as
\begin{equation}
\partial_t\omega + u\cdot\nabla\omega
= \nu\nabla^2\omega + (\nabla\times f)_z + (\nabla\times c(a_t))_z,
\qquad \nu=\mathrm{Re}^{-1}.
\end{equation}
The default flow forcing is $f_x=\sin(4y)$, $f_y=0$. The four-dimensional action controls additional sinusoidal forcing modes,
\begin{equation}
c_x(a_t)=\sum_{j=1}^{4}a_{t,j}\sin(k_j y),\qquad
(k_1,k_2,k_3,k_4)=(4,5,6,7),\qquad a_{t,j}\in[-0.5,0.5].
\end{equation}
The solver uses pseudo-spectral grids of $64\times64$ and $128\times128$ with two-thirds de-aliasing, timestep $\Delta t=10^{-3}$, and a control interval of $10$ simulation-time units. These settings follow the current HydroGym implementation~\citep{lagemann2026hydrogym}.

The repository-default RL observation is a flattened $8\times8$ grid of trajectory-averaged velocity magnitudes (64 scalars), providing a coarse, reduced representation of the flow. In contrast, we expose the instantaneous real-space vorticity field $s_t=\omega_t\in\mathbb{R}^{n\times n}$, with $n\in\{64,128\}$, at each control decision, thereby retaining the full spatial resolution of the simulated PDE state. The PPO MLP baseline receives the same vorticity field and flattens it before its policy network.

\paragraph{Reward.}
The current environment computes reward as $r_t=-\left(\alpha_E\,\overline{E}_t + \lambda_a\lVert a_t\rVert_1\right),$ where $\overline{E}_t$ is mean turbulent kinetic energy over the control interval and the released code fixes $\lambda_a=1$. We set $\alpha_E=-1$, matching the configuration used in our PPO runs, so the reward becomes $r_t=\overline{E}_t-\lVert a_t\rVert_1$. Alongside reward, we therefore report the mean executed $\ell_1$ action magnitude. This quantity measures the actuation term entering the reward directly and makes the reduction in control effort visible during training; reward itself still depends jointly on flow energy and actuation. 

\subsection{Foundation-model actor and baseline}
We evaluate \textbf{MORPH-Ti}, which has approximately 7M parameters and natively supports mixed-dimensional, variable-field PDE data~\citep{rautela2026morph}. For the results reported here, the pretrained MORPH backbone is frozen throughout PPO training; only the control heads and policy log standard deviation are optimized. The pretrained checkpoint is the general MORPH foundation-model checkpoint, with no prior fine-tuning on Kolmogorov control trajectories. More details about MORPH is given in Appendix~\ref{app:morph}.

As the conventional baseline, we use a full-field PPO MLP that receives the same real-space vorticity field and directly maps the flattened grid to a Gaussian policy. Across the compared conditions, the action space, reward, interaction budget, and PPO optimization protocol are held fixed. The corresponding settings are reported in Table~\ref{tab:setup}. All reported experiments are performed on a single NVIDIA A100 GPU with 40~GB of VRAM.

\begin{figure*}[h]
\vspace{-2mm}
\centering
\begin{minipage}{1.0\textwidth}
    \centering
    \includegraphics[width=0.85\linewidth, trim=0cm 0cm 15.5cm 0cm, clip]{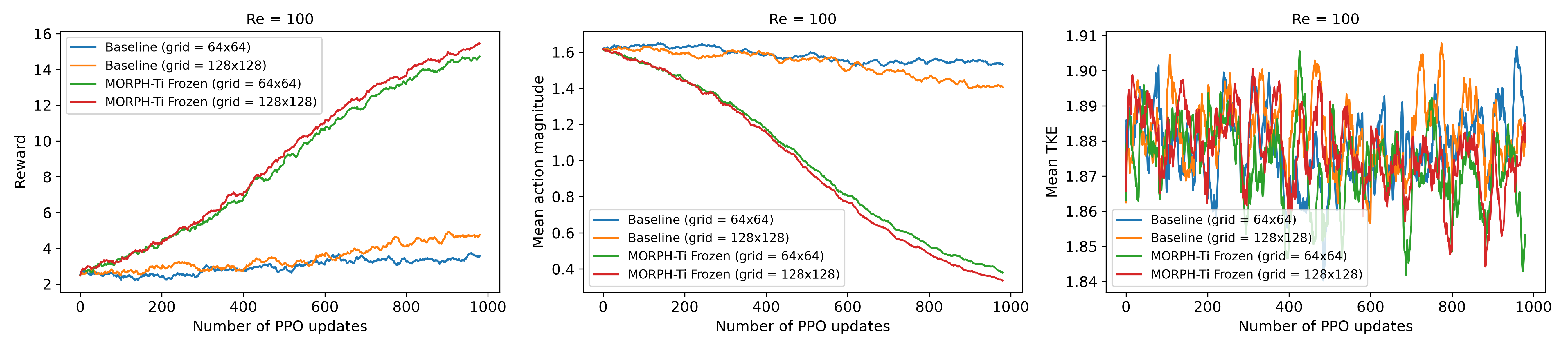}
\end{minipage}
\begin{minipage}{1.0\textwidth}
    \centering
    \includegraphics[width=0.85\linewidth, trim=0cm 0cm 15.5cm 0cm, clip]{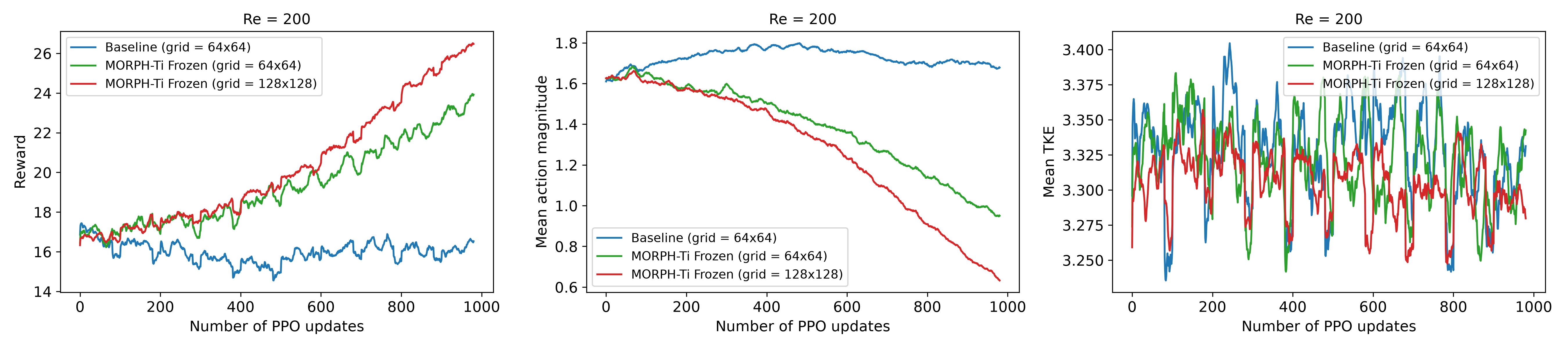}
\end{minipage}
\caption{Combined PPO training results across Reynolds number and spatial resolution. \textbf{Top:} $\mathrm{Re}=100$; \textbf{bottom:} $\mathrm{Re}=200$. \textbf{Left:} reward versus PPO update. \textbf{Right:} mean executed action magnitude, computed as the mean $\ell_1$ norm of the clipped four-dimensional control action. At $\mathrm{Re}=100$, baseline and frozen-MORPH results are shown for both $64\times64$ and $128\times128$ grids. Reward at each PPO update is the sum over its 10 rollout steps, with rewards
averaged across the eight parallel environments at each step. All curves are smoothed using a 50-update moving average for visualization.}
\label{fig:training_comparison}
\vspace{1em}
\begin{minipage}{1.0\textwidth}
\centering
\scriptsize
\setlength{\tabcolsep}{3.5pt}
\renewcommand{\arraystretch}{1.08}
\captionof{table}{\textbf{Model size and observed training wall-clock cost.}
Each PPO update collects $8\times10=80$ environment interactions, and throughput is computed as $80/t_{\mathrm{update}}$. The MORPH-Ti backbone is frozen; only the lightweight actor--critic components are optimized. Results are reported for $\mathrm{Re}=100$.}
\label{tab:compute}
\begin{tabular}{lcccc}
\toprule
Controller & Grid ($S_t$ resolution) & Trainable parameters &
Time/update (s) $\downarrow$ &
Throughput (interactions/s) $\uparrow$ \\
\midrule
Baseline PPO & $64\times64$ & 2.231M & $75 \pm 7$ & $1.067 \pm 0.100$ \\
Frozen MORPH-Ti & $64\times64$ & 0.134M & $152 \pm 5$ & $0.526 \pm 0.017$ \\
Baseline PPO & $128\times128$ & 8.522M & $86 \pm 9$ & $0.930 \pm 0.097$ \\
Frozen MORPH-Ti & $128\times128$ & 0.134M & $132 \pm 3$ & $0.606 \pm 0.014$ \\
\bottomrule
\end{tabular}
\end{minipage}
\end{figure*}

\subsection{Main results}
Fig.~\ref{fig:training_comparison} combines the PPO training results by Reynolds number and spatial resolution. Each horizontal axis reports PPO updates. We have used 8 parallel environments and 10 rollout steps, one update corresponds to 80 environment interactions and the 80,000-interaction budget yields 1000 updates.

At $\mathrm{Re}=100$, the frozen MORPH controller separates clearly from the full-field MLP baseline at both $64\times64$ and $128\times128$: its reward rises to roughly 15 by the end of training, while the baseline remains below 5. Simultaneously, the mean executed action magnitude decreases from about 1.6 to approximately $0.35$--$0.4$, whereas the baseline remains near $1.4$--$1.55$. At $\mathrm{Re}=200$, the $64\times64$ frozen controller reaches a reward near 24 compared with roughly 16 for the matched baseline, while the $128\times128$ controller reaches approximately 26. The corresponding action magnitudes decrease to roughly 0.95 and 0.65 for the frozen $64\times64$ and $128\times128$ controllers, respectively, compared with about 1.7 for the $64\times64$ baseline. Despite these differences in reward and actuation, TKE saturates at similar levels for both controllers: approximately $1.87$--$1.89$ at $\mathrm{Re}=100$ and $3.30$--$3.35$ at $\mathrm{Re}=200$. This indicates that the higher reward is not driven by a substantially different flow-energy state, but is instead associated primarily with a marked reduction in control effort. This behavior is qualitatively consistent with the Kolmogorov-flow results of \citet{lagemann2026hydrogym}, where effective policies achieve comparable TKE while requiring substantially less actuation. Overall, the frozen pretrained representation supports more control-efficient policies that maintain similar flow energetics while reducing the magnitude of the required forcing.

Table~\ref{tab:compute} highlights the parameter efficiency of the frozen MORPH controller, which trains only 0.134M parameters - $16.6\times$ and $63.4\times$ fewer than the MLP baseline at $64\times64$ and $128\times128$, respectively. Thus, MORPH requires substantially fewer task-specific trainable parameters while reusing its frozen pretrained representation. The reported wall-clock time for MORPH additionally reflects the overhead of coupling the PyTorch-based pretrained controller with the JAX-based HydroGym solver, including data conversion at the environment--policy interface.

\begin{figure*}[h]
\vspace{-2mm}
\centering
\begin{subfigure}{1.0\textwidth}
    \centering
    \includegraphics[width=0.9\linewidth, trim=0cm 0cm 0cm 1cm, clip]{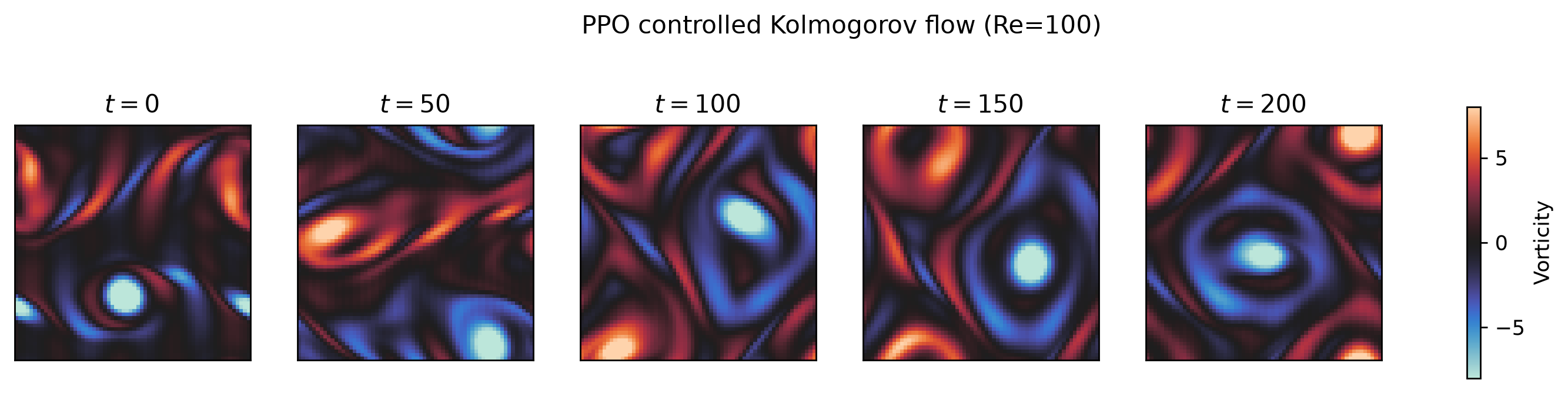}
    \caption{MORPH-Ti frozen rollout at $64\times64$ and $\mathrm{Re}=100$.}
\end{subfigure}
\begin{subfigure}{1.0\textwidth}
    \centering
    \includegraphics[width=0.9\linewidth, trim=0cm 0cm 0cm 1cm, clip]{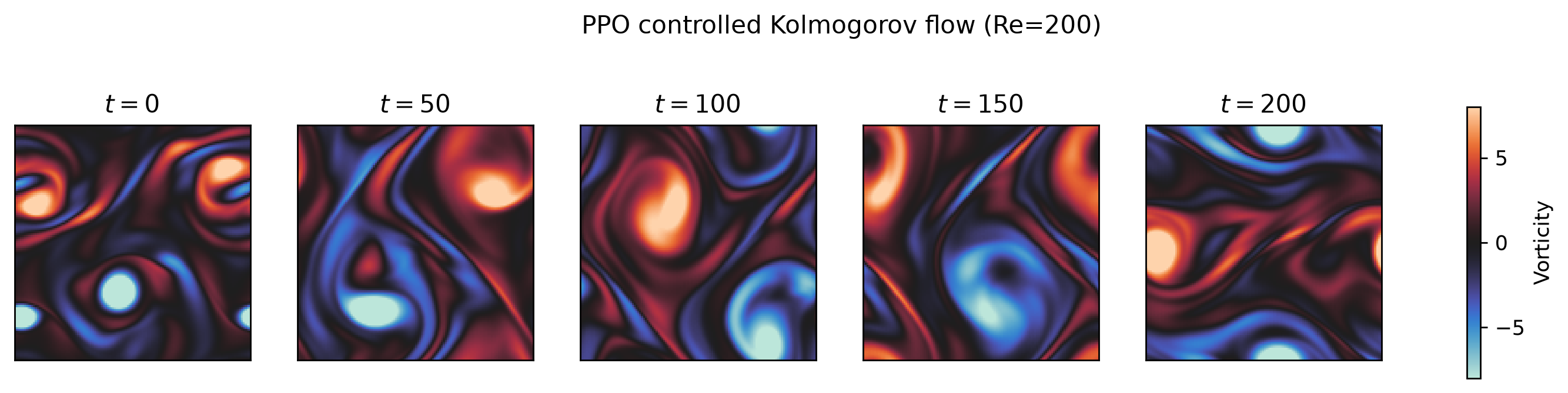}
    \caption{MORPH-Ti frozen rollout at $128\times128$ and $\mathrm{Re}=200$.}
\end{subfigure}
\caption{Closed-loop rollouts of the trained frozen MORPH-Ti controller. The snapshots show real-space vorticity at $t=0,50,100,150,200$. With a control interval of 10 simulation-time units, each trajectory spans 20 policy decisions.}
\label{fig:rollouts}
\vspace{-3mm}
\end{figure*}

Fig.~\ref{fig:rollouts} shows closed-loop rollouts of the trained frozen MORPH-Ti controller over 200 simulation-time units for both $64\times64$, $\mathrm{Re}=100$ and $128\times128$, $\mathrm{Re}=200$. The trajectories demonstrate that the learned policy can be deployed continuously in closed loop across different spatial resolutions and flow regimes while producing stable vorticity evolution over long horizon.

\section{Conclusion}
We repurpose a pretrained PDE foundation model for closed-loop flow control by replacing its predictive decoder with lightweight actor--critic heads. Across HydroGym Kolmogorov-flow experiments at $64\times64$ and $128\times128$ resolutions and $\mathrm{Re}=100,200$, the frozen MORPH-Ti controller achieves higher reward than matched full-field MLP baselines while using substantially smaller control actions. At the same time, it requires $16.6\times$ and $63.4\times$ fewer trainable parameters at $64\times64$ and $128\times128$, respectively. These results show that pretrained PDE representations can support effective and parameter-efficient state-to-action control without task-specific backbone adaptation. Rollouts further confirm stable long-closed-loop deployment across resolutions and flow regimes.

\section*{Acknowledgments}
Research presented in this article was supported by the Programmatic AI for LANSCE effort (internal charge code: CFLC) through the AOT-IC group at Los Alamos National Laboratory (LANL). This work used computational resources provided by the Darwin testbed at LANL, which is funded by the Computational Systems and Software Environments subprogram of LANL's Advanced Simulation and Computing (ASC) program (NNSA/DOE). Los Alamos National Laboratory is supported by the U.S. Department of Energy National Nuclear Security Administration under Contract No. 89233218CNA000001.

\bibliographystyle{plainnat}
\bibliography{references}

@article{rautela2026morph,
  title={Morph: Pde foundation models with arbitrary data modality},
  author={Rautela, Mahindra Singh and Most, Alexander and Mansingh, Siddharth and Love, Bradley C and Scheinker, Alexander and Oyen, Diane and Debardeleben, Nathan and Lawrence, Earl and Biswas, Ayan},
  journal={arXiv preprint arXiv:2509.21670},
  year={2025}
}

@article{ziegler2019fine,
  title={Fine-tuning language models from human preferences},
  author={Ziegler, Daniel M and Stiennon, Nisan and Wu, Jeffrey and Brown, Tom B and Radford, Alec and Amodei, Dario and Christiano, Paul and Irving, Geoffrey},
  journal={arXiv preprint arXiv:1909.08593},
  year={2019}
}

@inproceedings{saxena2026improved,
  title={Improved robustness of deep reinforcement learning for control of time-varying systems by bounded extremum seeking},
  author={Saxena, Shaifalee and Williams, Alan and Fierro, Rafael and Scheinker, Alexander},
  booktitle={2026 American Control Conference (ACC)},
  pages={1165--1172},
  year={2026},
  organization={IEEE}
}

@inproceedings{herde2024poseidon,
  title     = {Poseidon: Efficient Foundation Models for {PDE}s},
  author    = {Herde, Maximilian and Raoni{\'c}, Bogdan and Rohner, Tobias and K{\"a}ppeli, Roger and Molinaro, Roberto and de B{\'e}zenac, Emmanuel and Mishra, Siddhartha},
  booktitle = {Advances in Neural Information Processing Systems},
  volume    = {37},
  year      = {2024},
  doi       = {10.52202/079017-2311}
}

@misc{mccabe2025walrus,
  title        = {Walrus: A Cross-Domain Foundation Model for Continuum Dynamics},
  author       = {McCabe, Michael and Mukhopadhyay, Payel and Marwah, Tanya and Regaldo-Saint Blancard, Bruno and Rozet, Francois and Diaconu, Cristiana and Meyer, Lucas and Wong, Kaze W. K. and Sotoudeh, Hadi and Bietti, Alberto and Espejo, Irina and Fear, Rio and Golkar, Siavash and Hehir, Tom and Hirashima, Keiya and Krawezik, Geraud and Lanusse, Francois and Morel, Rudy and Ohana, Ruben and Parker, Liam and Pettee, Mariel and Shen, Jeff and Cho, Kyunghyun and Cranmer, Miles and Ho, Shirley},
  year         = {2025},
  eprint       = {2511.15684},
  archivePrefix= {arXiv},
  primaryClass = {cs.LG}
}

@inproceedings{hao2024dpot,
  title     = {{DPOT}: Auto-Regressive Denoising Operator Transformer for Large-Scale {PDE} Pre-Training},
  author    = {Hao, Zhongkai and Su, Chang and Liu, Songming and Berner, Julius and Ying, Chengyang and Su, Hang and Anandkumar, Anima and Song, Jian and Zhu, Jun},
  booktitle = {Proceedings of the 41st International Conference on Machine Learning},
  series    = {Proceedings of Machine Learning Research},
  volume    = {235},
  pages     = {17616--17635},
  year      = {2024},
  publisher = {PMLR}
}

@inproceedings{holzschuh2025pdetransformer,
  title     = {{PDE}-Transformer: Efficient and Versatile Transformers for Physics Simulations},
  author    = {Holzschuh, Benjamin and Liu, Qiang and Kohl, Georg and Thuerey, Nils},
  booktitle = {Proceedings of the 42nd International Conference on Machine Learning},
  series    = {Proceedings of Machine Learning Research},
  volume    = {267},
  pages     = {23562--23602},
  year      = {2025},
  publisher = {PMLR}
}

@article{lagemann2026hydrogym,
  title   = {The {HydroGym} reinforcement learning platform for fluid dynamics},
  author  = {Lagemann, Christian and Mokbel, Sajeda and Gondrum, Miro and R{\"u}ttgers, Mario and Wang, Yuning and Su{\'a}rez, Pol and Paehler, Ludger and Bezgin, Deniz A. and Buhendwa, Aaron B. and Callaham, Jared L. and Ahnert, Samuel and Zolman, Nicholas and Shao, Xiao and Loiseau, Jean-Christophe and Adams, Nikolaus A. and Meinke, Matthias and Schr{\"o}der, Wolfgang and Lagemann, Kai and Lagemann, Esther and Vinuesa, Ricardo and Brunton, Steven L.},
  journal = {Nature},
  year    = {2026},
  doi     = {10.1038/s41586-026-10917-6}
}

@article{rautela2026inverse,
  title={PDE foundation model-accelerated inverse estimation of system parameters in inertial confinement fusion},
  author={Rautela, Mahindra and Scheinker, Alexander and Love, Bradley and Oyen, Diane and DeBardeleben, Nathan and Lawrence, Earl and Biswas, Ayan},
  journal={arXiv preprint arXiv:2603.04606},
  year={2026}
}

@article{rautela2026out,
  title={Out-of-distribution transfer of PDE foundation models to material dynamics under extreme loading},
  author={Rautela, Mahindra and Most, Alexander and Mansingh, Siddharth and Pachalieva, Aleksandra and Love, Bradley and Malley, Daniel O and Scheinker, Alexander and Hickmann, Kyle and Oyen, Diane and Debardeleben, Nathan and others},
  journal={arXiv preprint arXiv:2603.04354},
  year={2026}
}

@inproceedings{chen2018neuralode,
  title     = {Neural Ordinary Differential Equations},
  author    = {Chen, Ricky T. Q. and Rubanova, Yulia and Bettencourt, Jesse and Duvenaud, David},
  booktitle = {Advances in Neural Information Processing Systems},
  volume    = {31},
  year      = {2018}
}

@article{lu2021deeponet,
  title   = {Learning nonlinear operators via {DeepONet} based on the universal approximation theorem of operators},
  author  = {Lu, Lu and Jin, Pengzhan and Pang, Guofei and Zhang, Zhongqiang and Karniadakis, George Em},
  journal = {Nature Machine Intelligence},
  volume  = {3},
  pages   = {218--229},
  year    = {2021},
  doi     = {10.1038/s42256-021-00302-5}
}

@article{ouyang2022training,
  title={Training language models to follow instructions with human feedback},
  author={Ouyang, Long and Wu, Jeffrey and Jiang, Xu and Almeida, Diogo and Wainwright, Carroll and Mishkin, Pamela and Zhang, Chong and Agarwal, Sandhini and Slama, Katarina and Ray, Alex and others},
  journal={Advances in neural information processing systems},
  volume={35},
  pages={27730--27744},
  year={2022}
}

@article{rautela2026sim+,
  title={Sim+Real: Joint Simulation-Experiment Training Improves Balanced Prediction in Physical Systems},
  author={Rautela, Mahindra and Scheinker, Alexander and Biswas, Ayan and Oyen, Diane and DeBardeleben, Nathan and Lawrence, Earl},
  journal={arXiv preprint arXiv:2610.01974},
  year={2026}
}

@article{li2020fno,
  title={Fourier neural operator for parametric partial differential equations},
  author={Li, Zongyi and Kovachki, Nikola and Azizzadenesheli, Kamyar and Liu, Burigede and Bhattacharya, Kaushik and Stuart, Andrew and Anandkumar, Anima},
  journal={arXiv preprint arXiv:2010.08895},
  year={2020}
}

@inproceedings{ronneberger2015unet,
  title     = {{U-Net}: Convolutional Networks for Biomedical Image Segmentation},
  author    = {Ronneberger, Olaf and Fischer, Philipp and Brox, Thomas},
  booktitle = {Medical Image Computing and Computer-Assisted Intervention},
  pages     = {234--241},
  year      = {2015},
  publisher = {Springer}
}

@article{rahman2022uno,
  title={U-no: U-shaped neural operators},
  author={Rahman, Md Ashiqur and Ross, Zachary E and Azizzadenesheli, Kamyar},
  journal={arXiv preprint arXiv:2204.11127},
  year={2022}
}

@article{maulik2021reduced,
  title   = {Reduced-order modeling of advection-dominated systems with recurrent neural networks and convolutional autoencoders},
  author  = {Maulik, Romit and Lusch, Bethany and Balaprakash, Prasanna},
  journal = {Physics of Fluids},
  volume  = {33},
  number  = {3},
  pages   = {037106},
  year    = {2021},
  doi     = {10.1063/5.0039986}
}

@article{solerarico2024beta,
  title   = {$\beta$-Variational autoencoders and transformers for reduced-order modelling of fluid flows},
  author  = {Solera-Rico, Alberto and Sanmiguel Vila, Carlos and G{\'o}mez-L{\'o}pez, Miguel and Wang, Yuning and Almashjary, Abdulrahman and Dawson, Scott T. M. and Vinuesa, Ricardo},
  journal = {Nature Communications},
  volume  = {15},
  pages   = {1361},
  year    = {2024},
  doi     = {10.1038/s41467-024-45578-4}
}

@article{schulman2017ppo,
  title={Proximal policy optimization algorithms},
  author={Schulman, John and Wolski, Filip and Dhariwal, Prafulla and Radford, Alec and Klimov, Oleg},
  journal={arXiv preprint arXiv:1707.06347},
  year={2017}
}

@article{rabault2019afc,
  title   = {Artificial neural networks trained through deep reinforcement learning discover control strategies for active flow control},
  author  = {Rabault, Jean and Kuchta, Miroslav and Jensen, Atle and R{\'e}glade, Ulysse and Cerardi, Nicolas},
  journal = {Journal of Fluid Mechanics},
  volume  = {865},
  pages   = {281--302},
  year    = {2019},
  doi     = {10.1017/jfm.2019.62}
}

@article{fan2020reinforcement,
  title   = {Reinforcement learning for bluff body active flow control in experiments and simulations},
  author  = {Fan, Dixia and Yang, Luning and Wang, Zhicheng and Triantafyllou, Michael S. and Karniadakis, George Em},
  journal = {Proceedings of the National Academy of Sciences},
  volume  = {117},
  number  = {42},
  pages   = {26091--26098},
  year    = {2020},
  doi     = {10.1073/pnas.2004939117}
}

@article{paris2021robust,
  title   = {Robust flow control and optimal sensor placement using deep reinforcement learning},
  author  = {Paris, Romain and Beneddine, Samir and Dandois, Julien},
  journal = {Journal of Fluid Mechanics},
  volume  = {913},
  pages   = {A25},
  year    = {2021},
  doi     = {10.1017/jfm.2020.1170}
}

@article{guastoni2023deep,
  title   = {Deep reinforcement learning for turbulent drag reduction in channel flows},
  author  = {Guastoni, Luca and Rabault, Jean and Schlatter, Philipp and Azizpour, Hossein and Vinuesa, Ricardo},
  journal = {The European Physical Journal E},
  volume  = {46},
  pages   = {27},
  year    = {2023},
  doi     = {10.1140/epje/s10189-023-00285-8}
}

@article{font2025deep,
  title   = {Deep reinforcement learning for active flow control in a turbulent separation bubble},
  author  = {Font, Bernat and Alc{\'a}ntara-{\'A}vila, Francisco and Rabault, Jean and Vinuesa, Ricardo and Lehmkuhl, Oriol},
  journal = {Nature Communications},
  volume  = {16},
  pages   = {1422},
  year    = {2025}
}

@article{suarez2025flow,
  title   = {Flow control of three-dimensional cylinders transitioning to turbulence via multi-agent reinforcement learning},
  author  = {Su{\'a}rez, Pol and Alc{\'a}ntara-{\'A}vila, Francisco and Rabault, Jean and Mir{\'o}, Arnau and Font, Bernat and Lehmkuhl, Oriol and Vinuesa, Ricardo},
  journal = {Communications Engineering},
  volume  = {4},
  pages   = {113},
  year    = {2025}
}

@article{montala2026wing,
  title   = {Flow separation control of an infinite wing section via multi-agent reinforcement learning},
  author  = {Montal\`a, Ricard and Font, Bernat and Su{\'a}rez, Pol and Rabault, Jean and Lehmkuhl, Oriol and Vinuesa, Ricardo and Rodriguez, Ivette},
  journal = {Journal of Fluid Mechanics},
  volume  = {1040},
  pages   = {A3},
  year    = {2026},
  doi     = {10.1017/jfm.2026.11837}
}

@article{mondal2026hypersonic,
  title   = {Hypersonic flow control: Generalized deep reinforcement learning for hypersonic intake unstart control under uncertainty},
  author  = {Mondal, Trishit and Jagtap, Ameya D.},
  journal = {Physics of Fluids},
  volume  = {38},
  pages   = {046112},
  year    = {2026},
  doi     = {10.1063/5.0324870}
}

\newpage
\appendix

\section{PPO details}\label{app:ppo}

\paragraph{Generalized advantage estimation.}
Advantages are computed using generalized advantage estimation (GAE). For each
transition, the temporal-difference residual is
\begin{equation}
    \delta_t
    =
    r_t
    + \gamma (1-d_t) V_{\psi_{\mathrm{old}}}(s_{t+1})
    - V_{\psi_{\mathrm{old}}}(s_t),
\end{equation}
where $d_t\in\{0,1\}$ denotes episode termination. The advantage estimate is
computed recursively as
\begin{equation}
    \hat{A}_t
    =
    \delta_t
    +
    \gamma\lambda(1-d_t)\hat{A}_{t+1},
\end{equation}
with $\gamma=0.99$ and $\lambda=0.985$. At the final step of each rollout,
$V_{\psi_{\mathrm{old}}}(s_{t+1})$ is obtained by bootstrapping from the final
observation. The corresponding value target is
\begin{equation}
    \hat{R}_t
    =
    \hat{A}_t + V_{\psi_{\mathrm{old}}}(s_t).
\end{equation}
Before evaluating the PPO policy objective, advantages are normalized within
each minibatch,
\begin{equation}
    \tilde{A}_t
    =
    \frac{\hat{A}_t-\mu_{\hat{A}}}
    {\sigma_{\hat{A}}+10^{-8}}.
\end{equation}

Accordingly, the clipped policy loss above uses $\tilde{A}_t$ in place of the
unnormalized advantage $\hat{A}_t$.

\paragraph{Value-function loss.}
The critic is optimized using the PPO clipped value objective. Let
$V_{\psi_{\mathrm{old}}}(s_t)$ denote the value prediction stored during
trajectory collection and $V_\psi(s_t)$ the updated prediction. We first define
the clipped prediction
\begin{equation}
    V_t^{\mathrm{clip}}
    =
    V_{\psi_{\mathrm{old}}}(s_t)
    +
    \operatorname{clip}
    \left(
        V_\psi(s_t)-V_{\psi_{\mathrm{old}}}(s_t),
        -\epsilon,\epsilon
    \right),
\end{equation}
where the same clipping parameter $\epsilon=0.2$ is used for the policy and
value objectives. The value loss is
\begin{equation}
    \mathcal{L}_{\mathrm{value}}
    =
    \frac{1}{2}
    \mathbb{E}_t
    \left[
    \max
    \left(
        \left(V_\psi(s_t)-\hat{R}_t\right)^2,
        \left(V_t^{\mathrm{clip}}-\hat{R}_t\right)^2
    \right)
    \right].
\end{equation}

\paragraph{Entropy regularization and total objective.}
The actor parameterizes a four-dimensional diagonal Gaussian policy. Its
entropy is
\begin{equation}
    \mathcal{H}(\pi_\theta)
    =
    \mathbb{E}_t
    \left[
        \mathcal{H}
        \left(
            \pi_\theta(\cdot\mid s_t)
        \right)
    \right],
\end{equation}
which encourages exploration by discouraging premature collapse of the action
distribution. The complete PPO objective minimized at each minibatch is
\begin{equation}
    \mathcal{L}_{\mathrm{PPO}}
    =
    \mathcal{L}_{\mathrm{policy}}
    +
    c_v \mathcal{L}_{\mathrm{value}}
    -
    c_e \mathcal{H}(\pi_\theta),
\end{equation}
with $c_v=0.5$ and $c_e=0.01$. Gradients are clipped to a maximum global norm
of $0.5$ before each optimizer step.

\paragraph{Numerical solver failures.}
As a safeguard against numerical breakdown of the PDE solver, any transition that produces a non-finite next state is treated as terminal \citep{saxena2026improved}. Specifically, for a failed transition we set
\begin{equation}
    r_t = r_{\mathrm{fail}}=-5,
    \qquad
    d_t = 1,
\end{equation}
and reset only the affected environment before continuing trajectory collection. Since GAE is multiplied by $(1-d_t)$, no value is bootstrapped across a solver failure or subsequent environment reset. Transitions for which the solver remains finite are unaffected by this safeguard.

\section{MORPH foundation model}\label{app:morph}
\subsection{Model choice}
We select MORPH-Ti as the foundation-model backbone because its pretraining interface and scale align naturally with the present control setting, not because other PDE foundation models are unsuitable. MORPH is pretrained with a one-step autoregressive objective, $X_t \mapsto X_{t+1}$, whereas models such as DPOT and Walrus condition prediction on an explicit temporal history. Our HydroGym controller instead follows the Markov state-to-action interface $\pi_\theta(a_t\mid s_t)$ and receives only the current vorticity field, so MORPH permits a direct prediction-to-action adaptation without introducing a history buffer or artificial temporal padding. Other foundation models, including Poseidon and PDE-Transformer, also avoid an explicit multi-step state history and could be adapted to control; MORPH is therefore used here as a controlled pilot testbed rather than as a uniquely capable architecture. We use the general MORPH checkpoint before any Kolmogorov-specific fine-tuning, and its compact $\sim$7M-parameter size makes frozen-backbone control experiments practical across multiple resolutions and flow regimes.

\begin{table}[h]
\centering
\caption{
Comparison of representative PDE foundation models for the present single-state control setting. The comparison concerns compatibility with state-to-action adaptation rather than forecasting performance. MORPH provides a compact backbone whose pretrained input and temporal interfaces can be retained without introducing an explicit history or lead-time variable.}
\label{tab:fm_selection}
\small
\setlength{\tabcolsep}{3.2pt}
\begin{tabular}{lccccc}
\toprule
\textbf{Model} &
\textbf{Pretraining input} &
\textbf{Scale} &
\textbf{History} &
\textbf{KF in pretraining} &
\textbf{Control adaptation} \\
\midrule
MORPH-Ti & $s_t$ & 7M & No & No & Direct \\
Poseidon-T & $(s_t,\Delta t)$ & 20.8M & No & No & Input/time adaptation \\
DPOT-Ti & $s_{t-9:t}$ & 7M & Yes & No & History required \\
Walrus & $s_{t-5:t}$ & 1.3B & Yes & No & History required \\
PDE-Transformer-S & $s_t$ & 33.2M & No & Yes & Direct \\
\bottomrule
\end{tabular}
\end{table}

\subsection{MORPH Architecture and Prior Applications}
MORPH is a modality-agnostic, autoregressive PDE foundation model designed to handle heterogeneous scientific data across 1D--3D domains, varying resolutions, and mixed scalar and vector fields \citep{rautela2026morph}. Its architecture combines component-wise convolutions for local field interactions, inter-field cross-attention for information exchange across physical variables, and factorized axial attention over spatial and temporal dimensions. A Unified Physics Tensor Format (UPTF-7) provides a common representation across heterogeneous datasets. The model is released open source in four sizes: Ti (7M), S (30M), M (126M), and L (480M) parameters. Beyond forward prediction, MORPH has been adapted to inverse parameter estimation in inertial confinement fusion using multimodal hyperspectral and scalar observations \citep{rautela2026inverse}, and evaluated for out-of-distribution transfer to shock-driven perturbed layered interfaces (PLI) and dynamic fracture/failure (FRAC) \citep{rautela2026out}. Recent work using MORPH demonstrates that joint simulation--experiment training achieves superior balanced performance across a broad range of evaluation weightings, while reducing simulation-domain forgetting and better preserving fields absent from experimental measurements \citep{rautela2026sim+}.

\end{document}